# Thinking Outside the Black Box: Insights from a Digital Exhibition in the Humanities

Sebastian Barzaghi[1], Alice Bordignon[2], Bianca Gualandi[3], Silvio Peroni[4]

[1] Department of Cultural Heritage, University of Bologna, Via Degli Ariani, 1, Ravenna, Italy - sebastian.barzaghi2@unibo.it

[2] Department of Classical Philology and Italian Studies, University of Bologna, Via Zamboni, 32, Bologna, Italy - alice.bordignon2@unibo.it

[3] Department of Classical Philology and Italian Studies, University of Bologna, Via Zamboni, 32, Bologna, Italy - bianca.gualandi4@unibo.it

[4] Department of Classical Philology and Italian Studies, University of Bologna, Via Zamboni, 32, Bologna, Italy - silvio.peroni@unibo.it

**ABSTRACT**[1]

One of the main goals of Open Science is to make research more reproducible. There is no consensus, however, on what exactly "reproducibility" is, as opposed for example to "replicability", and how it applies to different research fields. After a short review of the literature on reproducibility/replicability with a focus on the humanities, we describe how the creation of the digital twin of the temporary exhibition "The Other Renaissance" has been documented throughout, with different methods, but with constant attention to research transparency, openness and accountability. A careful documentation of the study design, data collection and analysis techniques helps reflect and make all possible influencing factors explicit, and is a fundamental tool for reliability and rigour and for opening the "black box" of research.

## 1. INTRODUCTION

In this contribution, we aim to anchor the discussion around open and reproducible research in the Arts and Humanities by presenting as a case study the creation of the digital twin of the temporary exhibition "The Other Renaissance: Ulisse Aldrovandi and the Wonders of the World"[2], currently under development within the PNRR Project CHANGES, and specifically its Spoke 4 – Virtual technologies for museums and art collections [1]. The original exhibition, held in Poggi Palace Museum (Bologna, Italy) between December 2022 and May 2023, consisted of more than 200 objects, mostly belonging to the naturalist Ulisse Aldrovandi and never exhibited before.

The creation of the digital twin – via the acquisition, processing, modelling, export, metadata creation, and upload of the 3D models to a web-based framework – was documented throughout in a structured manner in order to make the entire process transparent and reproducible. Indeed, no reproducibility is possible without transparency, or the careful and complete documentation of all relevant aspects of the study [6, p.5].

## 2. THEORETICAL BACKGROUND AND RELATED WORKS

Goodman and colleagues [6] suggest we should talk about three types of "reproducibility": (i) *methods reproducibility*, i.e. the ability to exactly reproduce a study by using the same raw data and the same methodologies to obtain the same results, (ii) *results reproducibility* – also referred to as *replicability* – i.e. the ability to obtain the same results from an independent study using the same methodologies as the original study, and (iii) *inferential reproducibility*, i.e. "the drawing of qualitatively similar conclusions from either an independent replication of a study or a reanalysis of the original study". In explaining how this differs from the two categories previously described, the authors add that scientists might "draw the same conclusions from different sets of studies and data or could draw different conclusions from the same original data, sometimes even if they agree on the analytical results" [6, p.4]. The reasons can be *a priori*, such as a different assessment of the probability of the hypothesis being explored, or can be linked to different choices about how to analyse and report data. This third type of reproducibility – which is also the most important according to the authors – might be the most common when talking about research reproducibility in the humanities. Peels and Bouter [12] look at how these concepts can be applied to the humanities, a field that has often been overlooked when talking about reproducible research. They prefer the terms "replicability" and "replication", and again they define three different levels: (i) *reanalysis*, that is Goodman et al.'s *methods reproducibility*, (ii) *direct replication*, where the same study protocol is applied to new data, and (iii) *conceptual replication*, where research data are new and the study protocol is modified [12]. These definitions do not perfectly overlap with those seen before but what is crucial to note here is that the authors find that, while replication can take various forms across the humanities, it is not fundamentally different from replication in the biomedical, natural, and social sciences and can be achieved by pre-registering the studies, and documenting and sharing methodologies and data

---

[1] All URLs mentioned in this paper have been last accessed on 05/04/2024.
[2] https://site.unibo.it/aldrovandi500/en/mostra-l-altro-rinascimento

[12]. Thanks to specific funding from the research funder NWO[3], a group of Dutch researchers conducted replication studies in a number of disciplines, including the humanities, and recently published a set of recommendations and lessons learned [3]. They found that in all cases replication studies help corroborate the findings of the original studies (e.g., extending the number of sources or using a more state-of-the-art approach) and can provide a more thorough understanding of the relevant research field and the available methodological choices [3, pp.6-7]. They also note, however, that "even experienced, highly conscientious researchers often find it difficult to document their protocols in enough detail to support direct replication" [3, p.8].

Those opposing the application of the reproducibility or replicability categories across all research areas cite the fact that, in several humanities domains, researchers may lack control over the experimental conditions of the original study, or have different viewpoints that produce different data and interpretations [9, pp.12-13; 13, p.7]. Carefully documenting the original study design, data collection and analysis, and reflecting on all possible influencing factors is fundamental for reliability and rigour but does not automatically ensure replicability [13, p.10]. Indeed, according to this view, to require replicability of all epistemic cultures is harmful and imposes "universal policies that fail to account for local (epistemic) differences", ultimately denying authority – and related rewards – to researchers in the humanities. On the other hand, the "umbrella of Open Science" is wide enough and its "accountability toolbox" is big enough to develop plural methods for assessing the quality of diverse research practices [13, p.12].

Leaving this discussion aside for now, in an increasingly open research environment, well-defined practices are essential for ensuring transparency, reliability, and equitable access to research outcomes. In the case of the digital twin of "The Other Renaissance" exhibition, we looked for some operational indications on how to achieve this goal in the literature produced in research fields relevant to the project. Wilson et al. [16] outline a set of recommendations for scientific computing, applicable across different disciplines and at varying levels of computational expertise. Regarding data management practices, their suggestions focus on the importance of incremental documentation and data cleaning. In particular, they advocate for continuous retention of raw data, robust backup strategies, data manipulation for improving machine and human readability and facilitating analysis, meticulous recording of the steps used to process data, using multiple tables in a way that each record in one table is interlinked with its respective representation in another table via a unique and persistent identifier, and using repositories that issue DOIs to the various data artefacts used and produced for easy access and citation. In the archaeological context, Karoune and Plomp [8] identify three distinct levels of workflow to make research activities reproducible, depending on the computational skills required to carry out such activities. Public access to research materials and methods is facilitated by the first level, which consists of transparent recording through documentation, requiring only the creation and maintenance of a written record of each analysis step, done in a format that allows other peers to read, comprehend, and replicate the work done, while requiring the least amount of computational expertise. Outputs at this level of workflow usually include documents describing the methods and processes, raw data files, and analysis output files. Ensuring version control through a shared file naming system and/or software with history tracking is also a common characteristic of transparent recording since it facilitates the documentation of the process as a whole.

### 3. MAKING THE DIGITISATION PROCESS MORE TRANSPARENT

To ensure a solid basis for transparency and replicability, our approach closely followed the aforementioned sets of best practices, in line with the indications listed in the Data Management Plan of the project [7]. The digitisation workflow involved creating two datasets as Google Sheet files shared between the team members: one (Object Table, or OT) for storing catalogue descriptions of the physical objects in the collection, the other (Process Table, or PT) for storing data about the digitisation process. After defining the structure of the tables, the variables represented by their headings, and the expected representation for each value, the data were populated in parallel by the team members. On the one hand, the OT was populated with data gleaned from official museum records and preliminary notes related to the exhibition objects, and thus was structured around a cataloguing description of each object (e.g. "title", "author", and so on). Where possible, controlled data values (e.g. people names, terms used for object types, etc.) were aligned with existing vocabularies (such as WikiData[4]) and authority lists (like VIAF[5] and ULAN[6]). On the other hand, the PT was populated with data inserted by the researchers during the acquisition of the objects and the creation of their 3D models and, thus, was structured around the steps involved in the overall digitisation process and their relevant attributes. Overall, the steps include an initial *acquisition activity* for capturing analogue materials and realising their preliminary digital representations, and a series of subsequent activities (*processing*, *modelling*, *export*, *metadata creation*, and *upload*) which involved the use of tools for refining and publishing the 3D models as usable, fully described scientific objects. In turn, these activities were represented as a set of information that included: the organisation responsible for the activity, the people responsible for actually carrying out the activity, the technique and/or tools used to perform the activity, and the timespan in which the activity was carried out. This preliminary work resulted in the creation of a record of the entire digitisation process. Google Sheets and

---

[3] https://www.nwo.nl/en/researchprogrammes/replication-studies
[4] https://www.wikidata.org/
[5] https://viaf.org/
[6] https://www.getty.edu/research/tools/vocabularies/ulan/

Microsoft Excel in the Microsoft Office 365 platform were strong facilitators for data retention, backup and versioning[7]. Moreover, shared formatting practices on elements such as dates and names were essential for preparing the data for the subsequent phases of the project. At the end of this stage, each object had its metadata, related digitisation phases with their features, and unique identifiers that allowed the two datasets to be linked to each other.

As information was added to both datasets, more work went into getting them ready to be published as machine-readable representations of the entire physical collection, its digital counterpart, and the procedure that, from the former, produced the latter. The Resource Description Framework (RDF)[8] was selected as a formal data representation for enabling transparent data publishing. However, in order to transform the current data into RDF statements, the table structures first had to be mapped to data models that could express and deepen the semantics of the data about cultural heritage and digitisation activities. We chose to reuse the CIDOC Conceptual Reference Model (CIDOC CRM)[9] [4] to represent the data detailing the physical and contextual attributes of the collection objects, and its extension CRM Digital (CRMdig)[10] [5] to depict the stages of the digitisation workflow. The Simplified Agile Methodology for Ontology Development (SAMOD) [14], a methodology to quickly create semantic models that are supported by rich documentation and test cases, was used to draw the needed conceptual constructs from CIDOC CRM and CRMdig and pack them into two data models.

## 4. MAKING INTERPRETATION MORE TRANSPARENT

One of the main goals of cultural heritage digitisation is the selection of specific elements of reality to store digitally. The selection process involves a deliberate human choice about the physical, geometric, chromatic, mechanical, and stylistic characteristics of the objects to digitise. These aspects are recorded inside a "grid of information", such as vectors, images, 3D models, databases, and tables, among others [2, p.127]. According to this logic, a digital technology survey is expected to approximate reality based on some predetermined features selected at the outset of the survey project. The quantity and quality of the data obtained during the survey significantly impact how accurate the digitisation will be. In this context, a *digital replica* is defined as an approximate, aesthetically convincing copy of a cultural site or artefact [2, p.127]. In our case study, the main aim was to obtain the digital version of the exhibition's experience, starting from the creation of its digital twin[11], linking to the digital assets of the various objects (3D and multimedia) in the collections, enriched by metadata, catalogued and accessible online using different devices [1, p.2].

Our approach for creating the digital twin of Aldrovandi's exhibition included in the first place the implementation of various setups and instruments to create morphologically precise models with highly detailed textures. Photogrammetry and structured light scanner (SLS) acquisition techniques have been used to obtain the digital representation of each item. The choice of these methodologies has been influenced by contextual factors (such as limited time and available space), materials, and the objects' size. We provided documentation about the challenges faced and related solutions adopted in the acquisition and processing phase. The documentation of the risks (e.g. acquisition of non-Lambertian materials, limited object's mobility, etc.) and the solutions adopted (e.g. cross polarisation techniques, specific setup schemas, etc.) permits others to retrace and repeat, at least in theory, the actions involved in a certain research effort, producing new data [15, p.2]. Concerning scanner acquisitions, we defined some common limits regarding texture final resolution, and we decided on a specific range for geometry complexity. During the entire process, open technologies and software were employed to maximise the workflow's re-adoption for the creation of a virtual exhibition in different settings. However, for some specific tasks (e.g. raw data elaboration), proprietary software was required since open-source software fails to produce satisfactory results.

Documenting processing decisions made for extra transparency should be a part of the scientific workflow and cultural heritage preservation. This can be done, as proposed by Moore et al. [10], by extracting a processing report from the photogrammetry software. Metashape[12] and 3DF Zephyr[13], the main software used for the photogrammetric processing phase, provide this option. The function has not been developed yet for the open-source alternative Meshroom[14], whose implementation in this project is under test. However, software for processing photogrammetric data is considered more open and transparent compared to software used for scanned data elaboration. Scanned data were elaborated using different versions of Artec Studio[15], which has proven to be a "black box" for those who do not own the software and the licence required to use it, allowing raw data export only in proprietary formats and without providing any processing report. Regarding modelling interventions, to guarantee transparency concerning the manipulation of the source data we provided different derivative versions for each 3D model. Level 0 represents the rough result obtained by the acquisition software, while level 1 includes the final high-definition model, where geometry issues have been fixed and lacking parts have been reconstructed. The comparison between these versions enables one to identify which parts were modelled and which parts belong to level 0. Level 2 instead includes the optimised model for web publication. Finally, we used as many standard and interoperable formats as possible for the generated data to facilitate their reuse on different platforms. Specifically, we used glTF, glb, obj, and mtl for 3D models; tiff, jpg, raw, and png for images; mp4 and mov for videos; and mp3 for audios.

---

[7] Since transparent recording does not involve any computational code, proprietary software like Google Sheets is acceptable as long as it includes features like versioning and exporting outputs to open formats (e.g., .txt, .rtf, .pdf) [7].

[8] https://www.w3.org/TR/rdf11-concepts/

[9] http://www.cidoc-crm.org/cidoc-crm/

[10] https://www.cidoc-crm.org/crmdig/; https://projects.ics.forth.gr/isl/CRMext/CRMdig_v3.2.2.rdfs

[11] Since cultural heritage may be intangible or temporary, Niccolucci et al. [11] suggest separating the data exchange dimension from the representation dimension for digital twins. This reconceptualisation rethinks data flows and bi-directionality as possible and as not mandatory requirements for digital twins of cultural heritage artefacts or landscapes, opening the possibility for accurate digital models (i.e. digital replicas) to evolve dynamically into a fully developed digital twin.

[12] https://www.agisoft.com/

[13] https://www.3dflow.net/it/

[14] https://alicevision.org/

[15] https://www.artec3d.com/it/3d-software/artec-studio

## 5. DISCUSSION AND CONCLUSIONS

We have described how the digitisation process of the exhibition "The Other Renaissance" has been documented throughout, with different methods, but with constant attention to research transparency, openness and accountability. Since any reality-capture or source-based model is affected by the lens of interpretation (of a human or software), tracking steps for the creation of a 3D model is essential to give transparency to these interpretations, facilitating the repeatability of the creation process [10]. Furthermore, data relating to the digitisation process can oftentimes be captured only once, while the process is ongoing, and it is therefore crucial to retain as much information as possible, structure it appropriately and make it available in an open and machine-readable format to provide a record of the entire physical collection, its digital counterpart, and the procedure that, from the former, produced the latter.

A particularly interesting aspect is the temporary nature of "The Other Renaissance" exhibition. At the time of writing, the exhibition concluded more than 6 months ago, objects on loan have been long returned, and the rooms where the exhibition took place have changed use. The same methodologies cannot be applied to the same data – Goodman et al.'s *methods reproducibility* or Peels and Bouter's *reanalysis* – because the physical collection does not exist in its original form anymore. What is possible, however, is that the methodologies described are applied to new data (different cultural heritage objects, exhibitions, etc.). Additionally, the careful documentation of the research process makes it possible for others to judge the relationship between the digital twin and the physical collection, a piece of information that is crucial for scientific scrutiny but that would otherwise have been irremediably lost on the day the temporary exhibition closed.

Documenting the project workflow in this manner is not simple: it requires careful planning, specific competencies, and it is extremely time-consuming. These efforts must be rewarded in the academic setting, if a culture of accountability, data curation and open, reproducible research is to become the norm. Initiatives like CoARA[16] are indeed nudging the scientific community in this direction, but while some practices – such as the publication of research data "as open as possible" and according to FAIR principles – are garnering increasing attention, the focus must be kept on methodologies, too, and on the need of carefully documenting each step of a research project. Further, as noted by Peels and Bouter [12], guidelines on how to report study protocols, methodologies and procedures are needed, and this is perhaps especially true in the humanities. The establishment of principles, like FAIR, and discipline-specific recommendations on how to manage and document research data in a transparent and traceable manner is a great first step in this direction. FAIR principles, supposedly discipline-agnostic, are being discussed and adapted to the different research cultures, Data Management Plans are becoming increasingly common, and templates and online tools are being produced to help researchers fill them out in a structured and machine-actionable manner. There is still more to be done, and more explicit attention needs to be devoted to research methodologies and how to document them in sufficient detail.

We recognise that the debate around the definition of reproducibility and replicability, and whether these terms should be applied to research as a whole, across all disciplines, is not settled [3;6;9;12;13]. However, there seems to be an agreement on the fact that research can be reproducible in varying degrees, from an "ideal" computational reproducibility all the way to fields where multiple interpretations of a certain phenomenon coexist. Replication here may help "filter out faulty reasoning or misguided interpretations, draw attention to unnoticed crucial differences in study methods" [12] but it is not always possible to ascertain which interpretation is correct. Circling back to the definition of *inferential reproducibility* [6] and the critique of the concept of replicability in a humanities context [9;13], the researchers' different viewpoints, theoretical background or previous assessments always have a bearing on how the study is conducted and how the results are interpreted. A careful documentation of the study design, data collection, and analysis techniques help reflect and make explicit all possible influencing factors, and is a fundamental tool for reliability and rigour and for opening the "black box" of research.

## ACKNOWLEDGEMENTS

This work has been partially funded by Project PE 0000020 CHANGES - CUP B53C22003780006, NRP Mission 4 Component 2 Investment 1.3, Funded by the European Union - NextGenerationEU.

## AUTHOR CONTRIBUTIONS STATEMENT

Authors' contribution according to CRediT (https://credit.niso.org/): Conceptualization (SB, AB, BG, SP); Investigation (SB, AB, BG); Supervision, Validation (SP); Writing – original draft, Writing – review & editing (SB, AB, BG).

---

[16] https://coara.eu/

**REFERENCES**

[1] Balzani, Roberto, Sebastian Barzaghi, Gabriele Bitelli, Federica Bonifazi, et al. 2024. «Saving Temporary Exhibitions in Virtual Environments: The Digital Renaissance of Ulisse Aldrovandi – Acquisition and Digitisation of Cultural Heritage Objects». *Digital Applications in Archaeology and Cultural Heritage* 32: e00309. https://doi.org/10.1016/j.daach.2023.e00309.

[2] Demetrescu, Emanuel, Enzo d'Annibale, Daniele Ferdani, e Bruno Fanini. 2020. «Digital Replica of Cultural Landscapes: An Experimental Reality-Based Workflow to Create Realistic, Interactive Open World Experiences». *Journal of Cultural Heritage* 41: 125–41. https://doi.org/10.1016/j.culher.2019.07.018.

[3] Derksen, Maarten, Stephanie Meirmans, Jonna Brenninkmeijer, Jeannette Pols, Annemarijn de Boer, Hans Van Eyghen, Surya Gayet, et al. 2024. «Replication Studies in the Netherlands: Lessons Learned and Recommendations for Funders, Publishers and Editors, and Universities». https://doi.org/10.31219/osf.io/bj8xz.

[4] Doerr, Martin, Christian-Emil Ore, e Stephen Stead. 2007. «The CIDOC Conceptual Reference Model - A New Standard for Knowledge Sharing», 51–56. https://doi.org/10.13140/2.1.1420.6400.

[5] Doerr, Martin, e Maria Theodoridou. s.d. «CRMdig: A Generic Digital Provenance Model for Scientific Observation».

[6] Goodman, Steven N., Daniele Fanelli, e John P. A. Ioannidis. 2016. «What does research reproducibility mean?» 8 (341).

[7] Gualandi, Bianca, e Silvio Peroni. 2024. «Data Management Plan: Second Version». https://doi.org/10.5281/ZENODO.10727879.

[8] Karoune, Emma, e Esther Plomp. 2022. «Removing Barriers to Reproducible Research in Archaeology». https://doi.org/10.5281/ZENODO.7320029.

[9] Leonelli, Sabina. 2018. «Rethinking Reproducibility as a Criterion for Research Quality». In Research in the History of Economic Thought and Methodology. Including a Symposium on Mary Morgan: Curiosity, Imagination, and Surprise, 36B:129–46. Emerald Publishing Limited. https://doi.org/10.1108/S0743-41542018000036B009.

[10] Moore, Jennifer, Adam Rountrey, e Hannah Scates Kettler. s.d. *3D Data Creation to Curation: Community Standards for 3D Data Preservation*. ALA. https://www.alastore.ala.org/content/3d-data-creation-curation-community-standards-3d-data-preservation.

[11] Niccolucci, Franco, Béatrice Markhoff, Maria Theodoridou, Achille Felicetti, e Sorin Hermon. 2023. «The Heritage Digital Twin: A Bicycle Made for Two. The Integration of Digital Methodologies into Cultural Heritage Research». *Open Research Europe* 3: 64. https://doi.org/10.12688/openreseurope.15496.1.

[12] Peels, Rik, e Lex Bouter. 2018. «The Possibility and Desirability of Replication in the Humanities». *Palgrave Communications* 4 (1): 95. https://doi.org/10.1057/s41599-018-0149-x.

[13] Penders, Holbrook, e De Rijcke. 2019. «Rinse and Repeat: Understanding the Value of Replication across Different Ways of Knowing». *Publications* 7 (3): 52. https://doi.org/10.3390/publications7030052.

[14] Peroni, Silvio. 2017. «A Simplified Agile Methodology for Ontology Development». In *OWL: Experiences and Directions – Reasoner Evaluation*, a cura di Mauro Dragoni, María Poveda-Villalón, e Ernesto Jimenez-Ruiz, 55–69. Lecture Notes in Computer Science. Cham: Springer International Publishing. https://doi.org/10.1007/978-3-319-54627-8_5.

[15] Rahal, Rima-Maria, Hanjo Hamann, Hilmar Brohmer, e Florian Pethig. 2022. «Sharing the Recipe: Reproducibility and Replicability in Research Across Disciplines». *Research Ideas and Outcomes* 8: e89980. https://doi.org/10.3897/rio.8.e89980.

[16] Wilson, Greg, Jennifer Bryan, Karen Cranston, Justin Kitzes, Lex Nederbragt, e Tracy K. Teal. 2017. «Good Enough Practices in Scientific Computing». *PLOS Computational Biology* 13 (6): e1005510. https://doi.org/10.1371/journal.pcbi.1005510.